\documentclass[journal]{IEEEtran}

\ifCLASSINFOpdf
\else
   \usepackage[dvips]{graphicx}
\fi
\usepackage{url,graphicx,amsmath,graphicx,commath,cite,enumitem,amssymb,multirow}
\usepackage{subcaption}
\usepackage{xcolor,hyperref,cite}

\begin{document}
% Title.
% ------
\title{TTS-Guided Training for Accent Conversion Without Parallel Data} % by Utilizing Text-To-Speech Pretraining}
%

% Single address.
% --------------
\author{Yi Zhou, Zhizheng Wu, Mingyang Zhang, Xiaohai Tian and Haizhou Li}

% \address{$^1$Department of Electrical and Computer Engineering, National University of Singapore, Singapore\\
% $^2$School of Data Science, The Chinese University of Hong Kong, Shenzhen, China}
%$^3$Bytedance AI lab, Singapore}

% \ninept
%
\maketitle
\begin{abstract}
Accent Conversion (AC) seeks to change the accent of speech from one (source) to another (target) while preserving the speech content and speaker identity. However, many AC approaches rely on source-target parallel speech data. We propose a novel accent conversion framework without the need of parallel data. Specifically, a text-to-speech (TTS) system is first pretrained with target-accented speech data. This TTS model and its hidden representations are expected to be associated only with the target accent. Then, a speech encoder is trained to convert the accent of the speech under the supervision of the pretrained TTS model. In doing so, the source-accented speech and its corresponding transcription are forwarded to the speech encoder and the pretrained TTS, respectively. The output of the speech encoder is optimized to be the same as the text embedding in the TTS system. At run-time, the speech encoder is combined with the pretrained TTS decoder to convert the source-accented speech toward the target. In the experiments, we converted English with two source accents (Chinese and Indian) to the target accent (American/British/Canadian). Both objective metrics and subjective listening tests successfully validate that, without any parallel data, the proposed approach generates speech samples that are close to the target accent with high speech quality. 

%In this paper, we propose a novel accent conversion framework that is guided by a pretrained text-to-speech (TTS) system without the need of parallel data. First, a TTS model is trained with large-scale speech data for the target accent. \textcolor{red}{(bad English - i don't understand) Then, a speech encoder is trained to convert the source-accented speech into a speech embedding with the target accent. } In particular, the pretrained TTS obtains a text embedding from the transcription of the source-accented speech. The training of the speech encoder aims to optimize the speech embedding to be in the same feature space as the text embedding. At run-time, the speech encoder is combined with the pretrained TTS decoder to convert the source-accented speech toward the target pronunciation. The experiments are conducted on two source accents, Chinese and Indian. Both objective metrics and subjective listening tests successfully validate that the proposed approach generates speech samples close to the target accent with a high speech quality. 

%\textcolor{red}{(to Zhou Yi: in the abstract, please talk about 1/ the motivation - what is the problem, and what is the challenges, and the idea behind the solutions (you missed these) 2/ briefly talk about what you plan to do, and argue why this is the best way; 3/ the conclusions.)}
\end{abstract}
\begin{figure*}
    \centering
    \includegraphics[width=.96\textwidth]{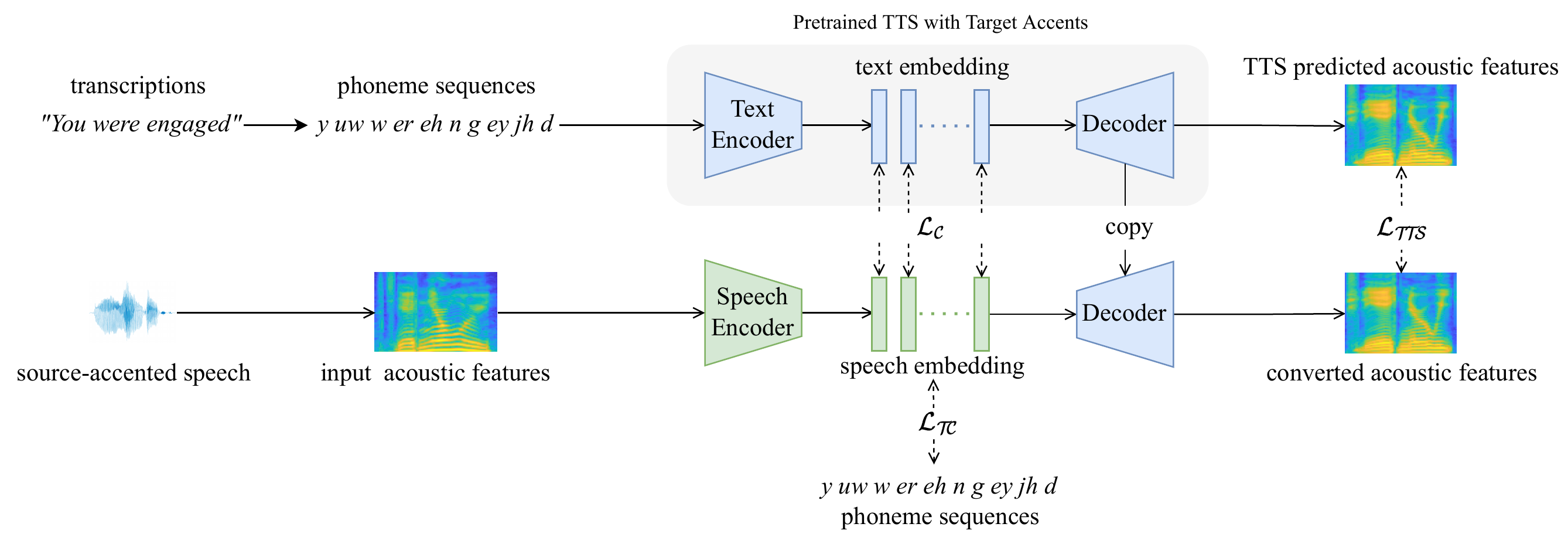}
    % \vspace{-0.3cm}
    \caption{The system diagram of the proposed AC framework. The TTS system is first pretrained with target-accented speech data. Then the speech encoder is trained with source-accented speech guided by the TTS system. The training objective is to minimize the combination of the contrastive loss $\mathcal{L}_{\text{C}}$ between the speech embedding and the text embedding, the text classification loss $\mathcal{L}_{TC}$ with the ground-truth phoneme sequences, and the reconstruction loss between the TTS predicted and converted acoustic features. Only the parameters in the speech encoder are updated.} % Training step 2 of the proposed FAC framework. $\mathcal{L}_{\text{C}}$ and $\mathcal{L}_{TC}$ denote the contrastive loss and text classification loss, $\mathcal{L}_{TTS}$ is the TTS guided loss. The text encoder and decoder form the TTS system. Step 2 training only updates the parameters of the speech encoder. The speaker embedding and vocoder are omitted for simplicity.} 
    \label{fig:system}
\end{figure*}

\section{Introduction}
\label{sec:intro}
Accent Conversion (AC) is a technique that converts speech with a source accent to a target accent. Ideally it is expected to modify only the accent-related speech attributes, while preserving the speech content and the speaker identity. Accent conversion technique will enable a variety of real-life applications, such as language education \cite{felps2009foreign,aryal2014can}, movie dubbing \cite{zhao2021converting}, and personalized TTS \cite{ding2022accentron}. %, AC draws increasing interest in recent research.

Among the many aspects of proficiency in speaking a language, e.g., lexical, syntactic, semantic, phonological, pronunciation is one of the most fundamental due to the neuro-musculatory basis of speech production \cite{scovel1988time,felps2009foreign,felps2012foreign}. An accent represents a distinctive way of pronouncing a language. There have been many studies to find the mapping between accents. Voice morphing methods perform accent conversion by decomposing and modifying the spectral details of speech~\cite{huckvale2007spoken,felps2010developing,aryal2013foreign}. Frame-wise feature mapping methods achieve the same by pairing source and target vectors based on their linguistic similarity ~\cite{aryal2014can}. Articulatory synthesizer methods take a different path by replacing the source pronunciation unit with the target counterpart through unit selection~\cite{felps2012foreign,aryal2015reduction,aryal2015articulatory}. All these studies rely on parallel speech data between the source and target accents, which limits the scope of applications~\cite{zhao2021converting,li2020improving}.

Another successful AC approach utilizes phonetic posteriorgram (PPG) to characterize the phonetic pronunciation \cite{zhao2018icassp,li2020improving} that doesn't require parallel speech data. It trains a speech synthesizer with the source-accented speech to map the PPGs to the acoustic features. Nonetheless, to synthesize the target accent for one speaker, a reference speech from another speaker with the target accent is still required.
Alternatively, the end-to-end framework can be trained on speech data with multiple accents, adding an additional accent encoder to the framework \cite{liu2020end, ding2022accentron}. In this way, speech with different accents can be obtained by varying the learned accent embedding without using reference speech. Recently, another reference-free AC framework with a pronunciation correction model has been explored \cite{zhao2021converting}. It first creates speech with the target accent in the desired speaker's voice. Then, the pronunciation correction model is trained to convert the source-accented speech to the synthesized target-accented speech.

In this paper, we propose a novel AC framework without parallel speech data. In the proposed framework, a TTS system is first trained with target-accented speech data. Then, a speech encoder is trained to convert the accent of the speech under the supervision of the pretrained TTS model. Specifically, source-accented speech and its corresponding transcription are respectively forwarded to the speech encoder and TTS system. The training of the speech encoder aims to minimize the distance between the its output and the hidden text embedding in the TTS system. In particular, the TTS system is trained with speech data containing the target accent only, and its hidden representations such as text embedding are assumed to be free of non-target accents. As a result, the output of the speech encoder is expected to be the same as the text embedding representing the target accent pronunciation.
During conversion, the speech encoder and the TTS decoder are combined for generating target-accented speech. It is worth mentioning that parallel speech data is not used in this work.

\section{Proposed Accent Conversion Framework}
\label{sec:proposed}
%The task of FAC can be formulated in two aspects: One is to create a feature space that represents the native accent, and the other is to find an effective mapping to convert the pronunciation of foreign accented speech into the native space. 
%The recently proposed approach \cite{zhao2021converting} successfully achieves FAC by addressing the two aspects. It first finds a native feature space with the synthesized speech in the target foreign speaker's voice. Then, a pronunciation correction model is implemented to convert the foreign accent speech to that of the native. 
The proposed AC framework utilizes a speech encoder to convert source-accented speech into the target pronunciation with the guidance of a pretrained TTS system. Next, we will introduce the training and conversion process in detail.

\subsection{Training Process}
\label{subsec:step1}
Among the various speech synthesis tasks, TTS systems, especially the end-to-end (E2E) models \cite{ping2017deep,shen2018natural,ren2020fastspeech}, have gained great success thanks to the power of deep learning and vast large-scale corpus contributed by the community \cite{huang2019voice}. The hidden text embedding in a well-trained TTS system is a robust representation underlying the speech content \cite{huang2021pretraining,zhang2021transfer}. If the TTS system is trained with target-accented speech only, the text embedding is expected to be free from other accents, which is highly desirable for AC.

Therefore, we propose to first train an E2E TTS system \cite{shen2018natural} with target-accented speech data from multiple speakers. The TTS system consists of a text encoder and a decoder, which can be found from Fig. \ref{fig:system}. The text encoder takes audio transcriptions as input and generates a text embedding, while the decoder generates acoustic features from the text representation conditioned on speaker embeddings \cite{ping2017deep,shen2018natural,ren2020fastspeech,zhang2021transfer}. 
%The total loss is the combination of the acoustic feature reconstruction loss and the stop token loss \cite{shen2018natural}. 
% We denote this E2E TTS as pretrained native TTS. 
It is assumed that the text embedding in this TTS system represents a feature space that characterizes pronunciation of the target accent.

Subsequently, we train a speech encoder to find an effective mapping to convert the pronunciation of source-accented speech into the target space. The speech encoder is an attention based network that uses acoustic features as inputs and predicts a speech embedding phoneme sequence. The attention in the encoder aligns the acoustic features and the phoneme sequences. The speech embedding is a bottleneck feature that represents the linguistic information before the phoneme prediction. Hence, the speech embedding has the equal length to the phoneme sequence as well as the text embedding, regardless of the speaking rate of speakers. 

Fig. \ref{fig:system} illustrates the proposed AC framework, where the pretrained TTS is fixed. The phoneme transcriptions of the source-accented speech are passed into the text encoder, and the acoustic features are used as the speech encoder input. The output text embeddings are taken by the speech encoder to learn the target pronunciation from acoustic features. The model parameters in the speech encoder are updated by the supervision of the corresponding phoneme sequence, text embedding, and the predicted acoustic features by the pretrained TTS. The total loss function is defined as follows:

\begin{equation}
\begin{aligned}
    \mathcal{L}_{SE} = w_{C}\mathcal{L}_{\text{C}} + w_{TC}\mathcal{L}_{TC} + \mathcal{L}_{TTS},
\end{aligned}
\label{eq:loss}
\end{equation}
where $\mathcal{L}_{SE}$ is the total loss of the speech encoder. $\mathcal{L}_{\text{C}}$ and $\mathcal{L}_{TC}$ denote the contrastive loss and text classification loss, while $w_{C}$ and $w_{TC}$ are their respective weights. The contrastive loss is used to guide the speech embedding toward the text embedding \cite{zhang2019non}. $\mathcal{L}_{TTS}$ is the reconstruction loss between the TTS predicted and the converted acoustic features \cite{shen2018natural}.

By completing the speech encoder training, the speech embedding should reside in the same feature space as the text embedding that represents the target accent information. The source-accented speech is projected onto the target phoneme representation. AC is thus achieved.

\subsection{Conversion Process}
During conversion, the speech transcriptions are not available. First, the speech encoder takes acoustic features extracted from the source-accented speech and encodes them into the speech embedding. Then, the decoder of the pretrained TTS generates acoustic features from the speech embedding. The converted speech waveform is obtained using a neural vocoder. 

\section{Experiments}
\label{sec:exp}
In this work, we refer to American, British, and Canadian English as target accents, while Indian and Chinese English are considered as source accents. We converted the English speech of Indian and Chinese speakers to the target accents.

\begin{table}
\caption{The description of parallel data requirements for different AC systems. The TTS system is omitted from this table since it is not an accent conversion system.}
\centering
\resizebox{.46\textwidth}{!}{\begin{tabular}{lccc}
\hline
\centering
% System &  parallel training\\ speech data &  reference native speech\\ during conversion \\
\multirow{2}{*}{system} & parallel speech data & reference speech \\
                        & during training      & during conversion       \\
\hline
(a) parallel-VC & Yes & No \\
(b) BNF-AC     & No  & Yes \\
(c) BNF-PC-AC  & Yes & No  \\
(e) \textbf{TTS-AC}     & \textbf{No}  & \textbf{No}    \\
\hline
\end{tabular}}
\label{table:data}
\end{table}

\subsection{Experimental Systems and Setups} 
In the experiment, we selected 4 speakers with the Indian or Chinese accents \textit{ASI} (male), \textit{TNI} (female), \textit{TXHC} (male), and \textit{LXC} (female) from the L2-ARCTIC corpus \cite{zhao2018l2}. For each speaker, 1031 utterances were used during training, and the remaining 100 utterances were used for conversion.  

The Parallel WaveGAN neural vocoder \cite{yamamoto2020parallel} was trained with VCTK corpus \cite{veaux2017cstr}. All speech signals were resampled at 16 kHz. The 80-dimensional Mel-spectrogram was extracted as the acoustic features with a window size of 50 ms and frameshift of 12.5 ms. 

We implemented 5 different systems using different techniques for comparison. Table \ref{table:data} gives an overview of the parallel data requirements of all conversion systems. We note that our proposed TTS-AC does not require parallel data during either training or conversion process.  

\begin{itemize}[leftmargin=*]
\item[]\textbf{(a) parallel-VC:} 
This is a sequence to sequence voice conversion (VC) system taking parallel speech data between two speakers with different accents. We selected speakers \textit{bdl} (male) and \textit{slt} (female) as the target, who were with the American accent from the CMU arctic database \cite{kominek2004cmu}. 
We had 4 source-to-target conversion speaker pairs: \textit{ASI-to-bdl}, \textit{TNI-to-slt}, \textit{TXHC-to-bdl}, and \textit{LXC-to-slt}. It converted the 80-dimensional Mel-spectrogram from the source-accented to the target-accented speech. The model followed the same implementation with the speech embedding to Mel-Spectrogram synthesizer in \cite{zhao2021converting}, and the only difference was the dimension of the input layer. In this system, converted speech does not preserve the speaker identity.

\item[] \textbf{(b) BNF-AC:} 
This is an AC system that takes PPG as input and generates acoustic features \cite{zhao2018icassp}. 4 speaker dependent models were trained with the source-accented speech of the selected speaker \cite{zhao2018icassp}. This system needed reference speech from speakers with the target accents for conversion. So, we used the same speech data from speaker \textit{bdl} and \textit{slt}, and there were 4 target-for-source speaker pairs: \textit{bdl-for-ASI}, \textit{slt-for-TNI}, \textit{bdl-for-TXHC}, and \textit{slt-for-LXC}. The 256-dimensional bottleneck features \cite{zhao2021converting} extracted by a pretrained acoustic model were used as PPG inputs. The bottleneck details can be found in \cite{zhou2021language}. The output was the 80-dimensional Mel-spectrogram. The model configurations were the same as the speech synthesizer in \cite{zhao2021converting}.

\item[] \textbf{(c) BNF-PC-AC:} 
This is an extension of system (b) with an additional pronunciation correction model. It first created target-accented speech in the desired speaker's voice using parallel speech data. Then, a pronunciation correction model was implemented to convert the acoustic features of source-accented speech to the synthesized target-accented speech \cite{zhao2021converting}. The pronunciation correction model used the same network configurations with the VC model in system (a). This system was speaker dependent, so we trained 4 models.

\item[] \textbf{(d) TTS:} This is a multi-speaker TTS system, which is the pretrained TTS described in Section \ref{subsec:step1}. It used speech data of American, British, and Canadian speakers from the VCTK corpus \cite{veaux2017cstr}. The selected training data included 63 speakers of approximately 20.5 hours. It adopted Tacotron2 network architecture \cite{shen2018natural} and the 256-dimensional speaker embedding \cite{g2e} was added to the decoder \cite{zhang2021transfer}. This serves as the upper bound of our proposed TTS-AC system.

\item[] \textbf{(e) TTS-AC:} 
This is the proposed AC system, as described in Section \ref{sec:proposed}. The TTS pretraining was the same as system (d). The speech encoder was speaker specific. The speech encoder consisted of 2 pyramid BLSTM layers \cite{chan2016listen} of 256 units in each direction, 1 LSTM layer of 512 units with location-aware attention, and a final softmax layer of 512 units. $w_{C}$ and $w_{TC}$ in the loss function \ref{eq:loss} were set to 30 and 1, respectively. At run-time, it adopted the beam search with a window size of 10.
\end{itemize}

\begin{table}
\centering
\caption{The average WER and PER results. source and target denote the natural speech with the source and target accents, respectively.}
\resizebox{.44\textwidth}{!}{\begin{tabular}{lccc}
\hline
\centering
% System & WER   & PER intra-accent  & PER inter-accent\\
\multirow{2}{*}{System} & \multirow{2}{*}{WER (\%)} & \multicolumn{2}{c}{PER (\%)} \\\cline{3-4}
\multicolumn{1}{l}{} &          & intra-accent  & inter-accent \\
\hline
(a) parallel-VC   & 18.14 & 50.36 & 56.94  \\
(b) BNF-AC       & 28.47 & 44.82 & 53.26 \\
(c) BNF-PC-AC    & 28.93 & 62.69 & 63.74 \\
(d) TTS           & 12.56 & 39.65 & 46.39  \\
(d) TTS-AC       & 15.75 & 40.37 & 48.52 \\\hline
source  & 29.83 & 52.22 & 55.26 \\
target  & \textbf{5.79} & \textbf{35.83} & \textbf{55.26} \\
\hline
\end{tabular}}
\label{table:objective}
\end{table}
\subsection{Objective Evaluations}
We report word error rate (WER), intra-accent and inter-accent phoneme error rates (PER) as objective metrics. 

\subsubsection{Word Error Rate (WER)}
Speech was transcribed using the Google Cloud Speech-to-Text\footnote{\url{https://cloud.google.com/speech-to-text}, accessed on 20/10/2022.} service. English (United States) was chosen, so the selected ASR was trained with a large amount of American English speech data. The results are summarized in Table \ref{table:objective}. A smaller WER value stands for a better performance. We note that the WER values of the source and target speech are 29.83\% and 5.79\%, respectively. The results tell the fact that ASR performance deteriorates with a different accent. Then, we further compare the generated speech from different systems. The WER results of system (b) and (c) are similar (28.47\% and 28.93\%), which are also comparable with the previous study \cite{zhao2021converting}. System (a) using parallel data achieves a better results of 18.14\%. 
The proposed TTS-AC system obtains a low WER value of 15.75\%, which is similar to the TTS system. It indicates that the proposed TTS-guided training greatly retains the capability of the pretrained TTS system in generating highly intelligible speech recognized by the ASR engine. 

\subsubsection{Phoneme Error Rate (PER)}
% Phoneme is defined as the smallest unit that discerns meaning between sounds. Accents have arisen from regions applying different phonemes (sounds) to graphemes (letters) when they pronounce words. 
Next, we report the intra-accent and inter-accent phoneme error rate (PER). The intra-accent PER is calculated between the converted speech and the natural speech with the target accent, while the inter-accent PER is the error between the converted speech and the natural speech with the source accent. The phoneme was extracted by the pretrained recognition encoder in \cite{zhang2019non} using target-accented speech data. 
%Ideally, the intra-accent PER should be relatively small while the inter-accent PER is large. 
The values are summarized in Table \ref{table:objective}. We can find that all inter-accent PER values are higher than the intra-accent, which means that the phonemes within the same accent are more similar than those between different accents. The target obtains the best 35.83\% and 55.26\% intra- and inter-accent PERs, the difference is 19.43\%. Whereas the source's is not so good, which may due to the reason that the phoneme recognizer degrades with non-target accents. 
The proposed TTS-AC system achieves a relatively low value for the intra-accent PER (40.37\%), and the value for the inter-accent PER increases to 48.52\%. 
% Besides, the BNF-FAC system also demonstrates a similar performance with the TTS-FAC system by obtaining comparable PER values. 
The results confirm that the proposed system with TTS supervision effectively facilities the AC network to generate speech with an accent that is close to the target in terms of recognized phonemes. 

\begin{figure}
    \centering
    \includegraphics[width=.46\textwidth]{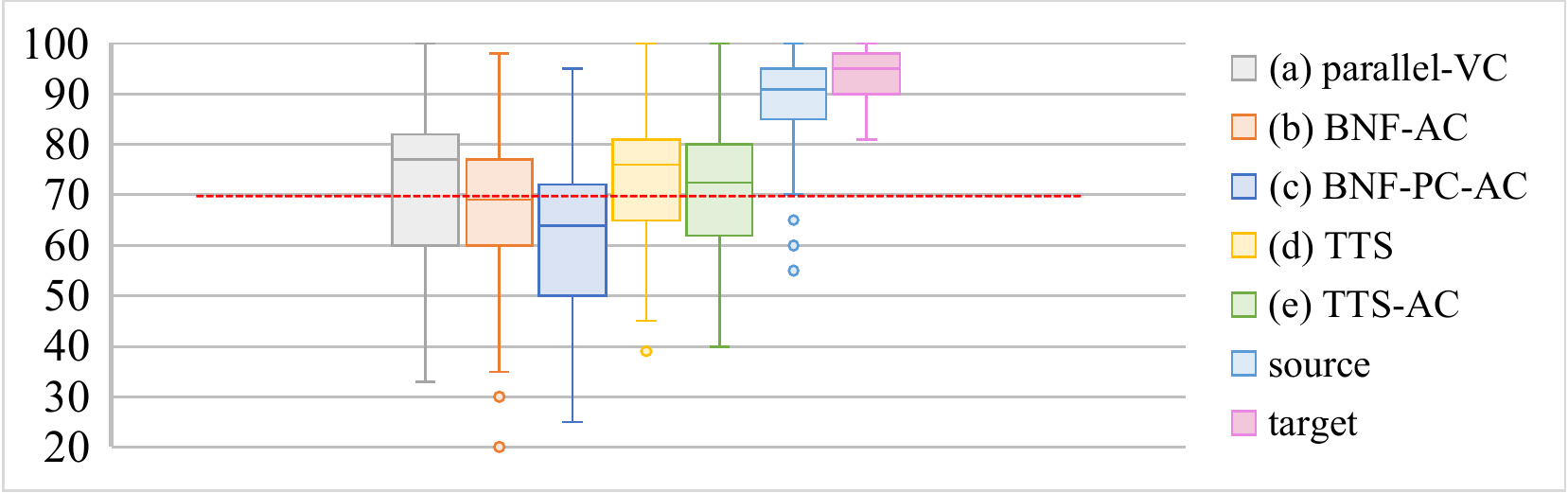}
    %   \vspace{-0.8cm}
    \caption{MUSHRA test results for speech quality. source and target refer to the natural speech with the source and target accents, respectively, which are used as reference.}
    \label{fig:mushra}
\end{figure}
\subsection{Subjective Evaluations}
We further conducted subjective listening tests using the Amazon Mechanical Turk platform\footnote{\url{https://www.mturk.com}}, including the MUSHRA test \cite{recommendation20011534}, the accentedness test, and the speaker similarity test. 25 American listeners participated in the tests and were paid upon completion of all tests. In each test, each system provides 10 utterances, which were randomly selected from 400 generated speech samples. Listeners were asked to compare samples of the same content from different systems in random order\footnote{Speech samples are at: \url{https://vcsamples.github.io/SPL2022AC/}}.

\begin{figure}
    \centering
    \includegraphics[width=.46\textwidth]{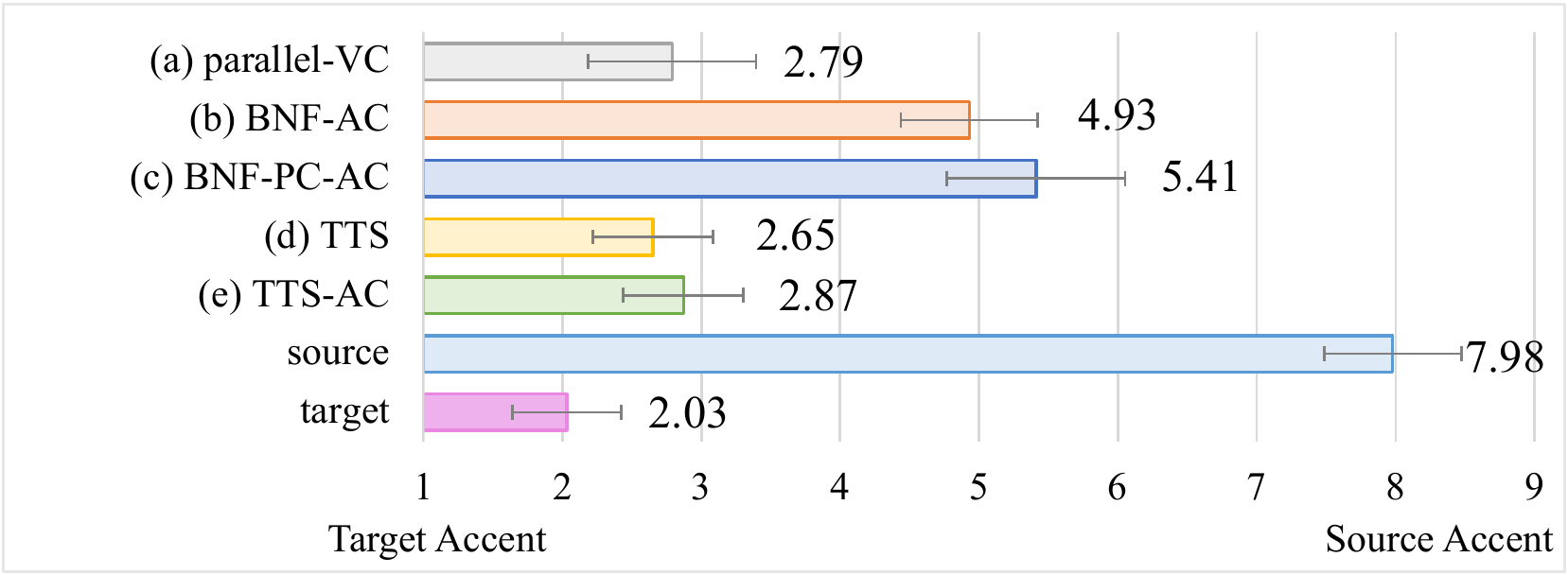}
    %   \vspace{-0.8cm}
    \caption{Accentedness test results with 95\% confidence intervals. A lower value means that the speech sounds more similar to the target accent, and vice-versa. The source and target are natural speech used as references.}
    \label{fig:accent}
\end{figure}

\subsubsection{MUSHRA Test for Speech Quality}
In this test, listeners were instructed to rate the speech quality of each sample between 0 and 100. A higher score corresponds to better speech quality. 
%The natural foreign accented speech of the target speaker was also included. 
%In total, one listener evaluated 50 samples by completing this test. 
Fig. \ref{fig:mushra} shows the MUSHRA score distributions. The proposed TTS-AC system performs slightly worse than the TTS system and similarly to the parallel-VC and the BNF-AC systems with an average MUSHRA score of over 70. Although the difference with the natural speech still exists, our proposed TTS-AC achieves a fairly effective good performance of generating high-quality speech waveform. 

\subsubsection{Accentedness Test}
This test assesses the speech accentedness. Listeners were instructed to rate the accentedness of an utterance on a nine-point Likert-scale (1: target accent; 9: source accent) \cite{munro1995foreign,zhao2021converting} given that any of the American, British, and Canadian accents are regarded as target accents.
%Each listener evaluated 60 samples in total. 
The listeners' choices are presented in Fig. \ref{fig:accent}. The natural speech with source and target accents receive ratings of 7.98 and 2.03, respectively. The accentedness rating of our proposed TTS-AC system is 2.87, far below the source-accented speech and very close to the target.
Moreover, it is encouraging to see that our proposed method without any parallel speech data achieves a comparable result to the parallel-VC system (2.79). Therefore, it can be concluded that our proposed AC technique with non-parallel training is very effective in converting the accent of speech samples from one to another. 

\begin{figure}
    \centering
    \includegraphics[width=.46\textwidth]{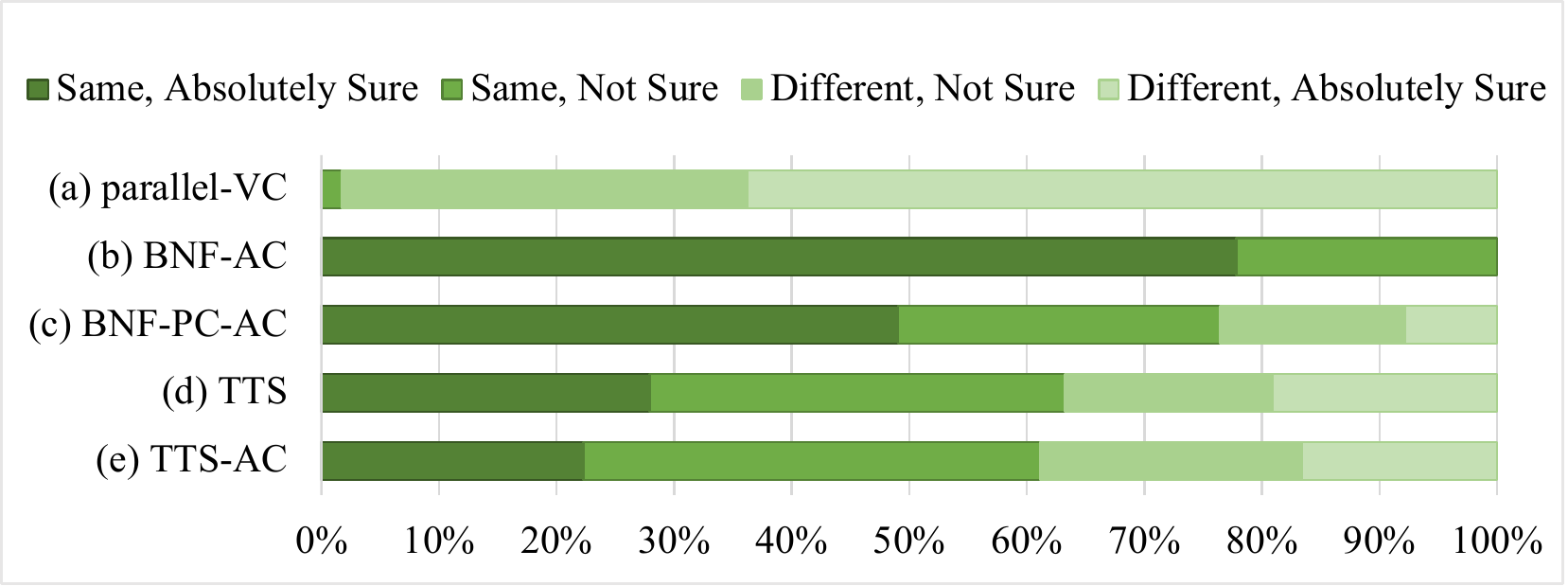}
    %   \vspace{-0.8cm}
    \caption{Speaker similarity test results. A higher percentage of `Same (not sure)' and `Same (sure)' together suggests a higher similarity to the desired target speaker.}
    \label{fig:vcc}
\end{figure}

\subsubsection{Speaker Similarity Test}
The speaker similarity test \cite{lorenzo2018voice} compared the speaker identity between the generated speech and the natural speech of the desired speaker. In this test, listeners had to give their decisions %over 30 samples 
on a scale of four \cite{lorenzo2018voice}. Fig. \ref{fig:vcc} depicts the results, where more than 60\% of the listeners thought that the converted speech of our TTS-AC system sounded the same as that of the desired speaker. Although this is a remarkably good score in terms of speaker similarity \cite{zhao2020voice}, we observe a performance drop in both TTS and TTS-AC systems. Since the proposed TTS-AC performs almost the same as the TTS system, we suspect that the speaker similarity problem lies in the speaker embedding, which is the only condition used to clone the voice in the TTS system. 
%It is suspected that the speaker embedding inaccuracies may be the cause of this performance degradation. 
%The inaccuracy in the speaker embedding may result in deviation of the generated voice. 
Similar findings are discussed in previous studies \cite{liu2020end, zhao2021converting}. 

\section{Conclusion}
\label{ssec:conclusion}
In this paper, we present an accent conversion technique without parallel data, where the conversion network training is guided by a pretrained text-to-speech (TTS) system. In the proposed framework, a TTS system is first trained with the target-accented speech data, and then a speech encoder is trained with the source-accented speech. The speech encoder learns the target pronunciation under the supervision of the text embedding obtained from the TTS system. The proposed method benefits from the robust hidden linguistic representation learnt from a pretrained TTS system. Therefore, the speech encoder can map source-accented speech to the target pronunciation without any parallel speech data. Experimental results successfully validate that our proposed method effectively converts the source to the target accent and generates high-quality speech. 
%In future, we will explore other pronunciation mapping techniques and try to address the generated speaker similarity.

%greatly reduces the foreign accent and performs close to the voice conversion with parallel speech data. 
%In future, we will address the generated speaker identity and explore other pronunciation mapping mechanism.

% \section{Acknowledgement}
% \label{sec:ack}
% DSO project
\bibliographystyle{IEEEtr}
% \bibliography{strings,refs}

\end{document}